\documentclass{ws-procs10x7}
\usepackage{balance}

\newcolumntype{d}[1]{D{.}{.}{#1}}

\def\be{\begin{eqnarray}}
\def\en{\end{eqnarray}}

\makeindex
\begin{document}

\title{Search for Bottom Counterparts of $X(3872)$ and $Y(4260)$
via $\pi^+\pi^-\Upsilon$}

\author{George W.S. Hou$^*$}

\address{Department of Physics, National Taiwan
University, Taipei, Taiwan 10617, R.O.C.\\
$^*$E-mail: wshou@phys.ntu.edu.tw}


\twocolumn[\maketitle\abstract{The $X(3872)$ and $Y(4260)$, both
discovered in $\pi^+\pi^- J/\psi$ mode, are rather unusual:
$X_c\equiv X(3872)$ is very narrow, while $Y_c \equiv Y(4260)$ has
large $Y_c\to \pi^+\pi^-J/\psi$ width. Many models for their
composition have been suggested, but perhaps discovering their
bottom counterparts could shed much light on the issue. The narrow
state, $X_b$ may be searched for at the Tevatron via $p\bar p \to
\pi^+\pi^-\Upsilon + X$, with the LHC much more promising. The
state $Y_b$ can be searched for at B factories via radiative
return $e^+e^- \to \gamma_{\rm ISR} + \pi^+\pi^-\Upsilon$ on
$\Upsilon(5S)$, or by $e^+e^- \to \pi^+\pi^-\Upsilon$ direct
scan.}
 ]

\section{INTRODUCTION}


After laying dormant for two decades, charmonium physics is
experiencing a renaissance lately, largely due to the
unprecedented luminosities achieved at the B factories. We now
have\cite{Pakhlov} an $X$ and a $Y$ and a $Z$ of 3940 MeV states
produced via various mechanisms. But two states stand out
especially: the $X(3872)$ and the $Y(4260)$, both observed in the
$\pi^+\pi^- J/\psi$ channel.

The $X_c \equiv X(3872)$ was discovered\cite{X3872} by Belle. A
very narrow state with $\Gamma < 2.3$ MeV (consistent with
experimental resolution) was observed in $\pi^+\pi^- J/\psi$
recoiling against $K^+$ from $B^+$ decay. Owing to this
narrowness, it was quickly confirmed by CDF\cite{XCDF} and D0 in
$p\bar p \to \pi^+\pi^-
J/\psi + X$. 
The mass of $X_c$ is just at the $D^0\bar D^{*0}$ threshold, which
may be related to its narrowness. Theoretical interpretations for
$X_c$ has ranged from $D^0\bar D^{*0}$ molecule, 4 quark state, to
charmonium hybrid.

The $Y_c\equiv Y(4260)$ was discovered\cite{Y4260} by BaBar in
initial state radiation (ISR, or ``radiative return") $e^+e^- \to
\gamma_{\rm ISR} + \pi^+\pi^- J/\psi$ events, hence has J$^{\rm
PC} = 1^{--}$. It has a normal hadronic width of $\sim 90$~MeV.
However, $\Gamma_{ee}\,{\cal B}_{\pi^+\pi^-J/\psi} \sim 5.5$ eV
implies (cf. $\sim 0.5$ eV for $\psi(3770)$) a rather large $Y_c
\to \pi^+\pi^- J/\psi$ rate.
These properties were confirmed\cite{CLEOc} by CLEO-c via $e^+e^-
\to \pi^+\pi^- J/\psi$ energy scan.
Theoretical interpretations for $Y_c$ range from hybrid, 4 quark
state, meson molecule or baryonium, to conventional $\psi(4S)$.

So, what about bottomonium? Clearly, the analogous $X_b$, $Y_b$
states should be searched for in the $\pi^+\pi^- \Upsilon$
channel.\cite{Hou}
We point out that the narrow state $X_b$ can be searched for at
the Tevatron and the LHC. The $1^{--}$ state $Y_b$ can be searched
for at the B factories, and future Super B factory, either by ISR
search on the $\Upsilon(5S)$, or by direct scan at $\Upsilon(5S)$
energies and beyond.

\section{Tevatron/LHC: $X_b \to \pi^+\pi^- \Upsilon$}

What is the $X_b$ mass? It could be at the $BB^*$ threshold of
10640 MeV, but it could also be lower, e.g. 10560 MeV, by coupled
channel arguments. The mass itself is to be probed to shed light
on the hadronic structure.

For $X_b$ production, we no longer have the analogy of $B\to KX$,
and since the preferred J$^{\rm PC}$ is $1^{++}$, we have to
resort to hadronic machines. Fortunately, prompt $X(3872)$
production, similar to $\psi'$, dominates.

The current projection\cite{Kerzel} is for roughly 3000
reconstructed $X(3872)$ per fb$^{-1}$. From 40 pb$^{-1}$ data one
can\cite{psiCDF} reconstruct $\sim 3\times 10^5$ $J/\psi$. In
comparison, from 77 pb$^{-1}$ data, $\sim 4400$ $\Upsilon$ is
reconstructed.\cite{UpsCDF} The leptonic branching fraction has
been taken into account. We estimate that  $\sim 180$
reconstructed $X_b \to \pi^+\pi^- \Upsilon$ events may be possible
for 8 fb$^{-1}$. More conservatively, we take the ratio of the
peak cross sections for reconstructed $\Upsilon$ and $J/\psi \to
\mu\mu$ events, which is $\sim 1/800$, and reach an estimate of
$\sim 30$ $X_b \to \pi^+\pi^- \Upsilon (\to \mu\mu)$ events.
However, the relative BRs and production fractions for $X_b$ vs
$X_c$ are unknown, which is in fact to be probed in the search. We
can only conclude that $X_b$ might be accessible.

What needs to be done at the Tevatron Run II is to benchmark the
$\Upsilon(2S) \to \pi^+\pi^- \Upsilon$ process, then look for a
higher mass ``bump", much like the case for $X(3872)$.\cite{XCDF}

The LHC is much more promising. For $\Upsilon$ production, a
study\cite{Domen} using PYTHIA fits to the results of $\Upsilon$
production at the Tevatron, then extends to the LHC. The peak
cross section increases by over two orders of magnitude, and
${\cal B} (\Upsilon \to \mu^+\mu^-) \, d\sigma(pp\to \Upsilon +
X)/dp_T$ at LHC is roughly 1/10 that of ${\cal B} (J/\psi \to
\mu^+\mu^-) \, d\sigma(p\bar p\to J/\psi + X)/dp_T$ at the
Tevatron. Thus, one expects over 1000 $X_b \to \pi^+\pi^- \Upsilon
(\to \mu\mu)$ events per fb$^{-1}$, and discovery at the LHC seems
assured, if the $X_b$ exists.

The homework for ATLAS, CMS and LHCb is the same as at the
Tevatron: benchmark $\Upsilon(2S) \to \pi^+\pi^- \Upsilon$, then
look for higher mass narrow state(s). It is highly desirable to
improve cross section calculations.

Note that the forward production of $b\bar b$ pairs should be
enhanced, although even less is known compared to central
production. To take advantage of this, and because of the interest
in heavy flavor physics and CP studies, the dedicated LHCb
experiment has forward design with the luxury of RICH and ECAL
detectors, i.e. much better hadron identification and photon
detection than ATLAS and CMS. Thus, though discovery of $X_b$ may
be an open competition between the three LHC experiments, LHCb may
have the best capability in exploiting this new bottomonium
spectroscopy, such as narrow states in $\omega \Upsilon$ and
$K^+K^-\Upsilon$. Of course, charmonium spectroscopy can also be
studied.


\section{$Y_b \to \pi^+\pi^- \Upsilon$ in $e^+e^- \to (\gamma)Y_b$}

With 211 fb$^{-1}$ data, BaBar discovered\cite{Y4260} the
$Y(4260)$ in radiative return $e^+e^- \to \gamma_{\rm ISR}+
\pi^+\pi^-J/\psi$. From 125 reconstructed $Y(4260)\to
\pi^+\pi^-J/\psi$ events,
\begin{equation}
\Gamma_{Y_c\to ee}{\cal B}_{Y_c\to \pi^+\pi^-J/\psi} \sim 5.5\
{\rm eV},
\end{equation}
is inferred, while $\Gamma_{Y_c} \simeq 88$ MeV. This gives ${\cal
B}_{Y_c\to ee}{\cal B}_{Y_c\to \pi^+\pi^-J/\psi} \sim 7\times
10^{-8}$, which is larger than the case for $\psi(4040)$ and
$\psi(4160)$. But since $Y_c(4260)$ falls at a dip in the cross
section for $e^+e^-\to$ hadrons, barring subtle interference
effects, presumably $\Gamma_{Y_c\to ee} \ll \Gamma_{\psi(4160)\to
ee} \sim$ 770 eV. Hence, the partial width $\Gamma_{Y_c\to
\pi^+\pi^-J/\psi}$ should be a few MeV or higher, much larger than
typical charmonia.

Stimulated by this, CLEO-c made a scan\cite{CLEOc} for $\sqrt s =
3.97$ to 4.26 GeV, which covers the $\psi(4040)$,  $\psi(4160)$
and $Y_c(4260)$, and 15 decay channels were studied. With just
0.013 fb$^{-1}$ around 4260 MeV, CLEO-c confirmed the BaBar
signal, finding $\sigma(e^+e^- \to \pi^+\pi^-J/\psi) \simeq 58$
pb, which is consistent with Eq.~(1). CLEO-c also found signals in
$\pi^0\pi^0J/\psi$ and $K^+K^-J/\psi$ channels.

What is noteworthy is that, with accumulated data that is only
1/16000 that of the ISR data, CLEO-c reconstructed 35
$\pi^+\pi^-J/\psi$ events with low background, while the ISR
approach gave 125 events after accounting for background. For
$Y_b$ search, therefore, one should consider both ISR return  and
direct scan, i.e. try both $e^+e^- \to \gamma_{\rm ISR}
\pi^+\pi^-\Upsilon$ {\it and} $e^+e^- \to \pi^+\pi^-\Upsilon$
processes, and weigh the benefits against cost. It should be clear
that a $\sim 100$ MeV width precludes hadronic machines as an
effective search tool because of high background.

It is interesting to note that, motivated by $B_s$ physics, Belle
has now accumulated\cite{Drutskoy} $\sim 24$ fb$^{-1}$ data on the
$\Upsilon(5S)$. The study not only demonstrated the capability to
accumulate $\sim 1$ fb$^{-1}$ per day on $\Upsilon(5S)$, a scan of
5 energy points of 0.03 fb$^{-1}$ each around the $\Upsilon(5S)$,
altogether done within a day, demonstrates the relative ease in
changing energies. So, one not only has significant amount of data
at hand to probe for $Y_b$ lighter than 10870 MeV, direct scan can
be contemplated.

So what is the target mass range for $Y_b$? And what cross
sections can one expect for the two advocated processes? We can
only make guestimates, using the CLEO-c favored\cite{CLEOc} $Q\bar
Qg$ hybrid picture as a guide.

Lattice studies have put the lowest $b\bar bg$ hybrid at
around\cite{Michael} 10700--11000 MeV. The $1^{--}$ quantum number
is possible, but many other quantum numbers are possible,
including exotic ones such as $1^{-+}$. The $1^{--}$ can mix with
standard $s$-wave mesons and may not be the lightest, but it is
clearly the most accessible.
Lattice studies tend to give lightest $c\bar cg$ hybrid mass
around 4400 MeV. If $Y_c(4260)$ is indeed dominantly a hybrid, by
analogy the lattice range for the $b\bar bg$ hybrid should be
scaled down to 10600--10900 MeV. This would make $\Upsilon(5S)$,
at 10865 MeV, an excellent place to conduct $e^+e^- \to
\gamma_{\rm ISR}Y_b\to \gamma_{\rm ISR}\, \pi^+\pi^-\Upsilon$
search. We take 10600, 10700 and 10800 MeV as nominal $M_{Y_b}$
values for purpose of illustration.

\begin{table}[t]
\tbl{Cross section for radiative return from $\Upsilon(5S)$, and
for direct $e^+e^-\to \pi^+\pi^-\Upsilon$ scan, for $M_{Y_b} =$
10600, 10700 and 10800 MeV. The branching ratio product is taken
as the same as Eq. (1).
 }
{\begin{tabular}{|c|c|c|c|} \hline
\raisebox{0pt}[10pt][6pt] Process
 & 10600 & 10700 & 10800   \\ \hline
\raisebox{0pt}[10pt][6pt] {$e^+e^-\,\to\,\gamma_{\rm ISR}
\pi^+\pi^-\Upsilon$}
 & $0.4\,$pb & $0.6\,$pb & $1.6\,$pb  \\ \hline
\raisebox{0pt}[10pt][6pt] {$e^+e^-\,\to\,\pi^+\pi^-\Upsilon$}
 & $9.1\,$pb & $9.0\,$pb & $8.8\,$pb \\ \hline
\end{tabular}}
\end{table}

We caution, however, that even with lattice studies of hybrids,
there are uncertainties due to difference in numerical approach,
scale uncertainty, as well as treatment of dynamic quarks. For
example, some studies\cite{Kuti} find the lowest $b\bar bg$ hybrid
mass to be $\sim 10900$--11000 MeV, while giving the right mass
for $c\bar cg$ hybrid that is consistent with $Y_c(4260)$. If
$Y_b$ is heavier than 10900 MeV, then it cannot be accessed by
$\Upsilon(5S)$ data, and a direct scan would be necessary.

For other properties of $Y_b$, a width of order 100 MeV seems
reasonable, while the product branching fraction similar to
Eq.~(1) can also be assumed, although $\Gamma_{Y_b\to
\ell\ell}{\cal B}_{Y_b\to \pi^+\pi^-J/\psi}$ could be smaller or
larger than Eq.~(1).

The ISR cross section is well known. In the narrow width
approximation, and leading order in $\alpha$, one
has\cite{Eidelman}
\begin{eqnarray}
\sigma_{\rm ISR} \simeq 36 \frac{\Gamma_{ee}{\cal B}_{
\pi^+\pi^-\Upsilon}}{M_{Y_b}}
\left(\frac{1}{x}-1+\frac{x}{2}\right) \mu{\rm b}, 
\end{eqnarray}
where $x = 1 - M_{Y_b}^2/s$ is the energy fraction carried away by
the ISR photon (usually not observed) in the CM frame. The cross
sections for our representative values of $M_{Y_b} = 10600$, 10700
and 10800 MeV are given in Table~1.

Radiative return cross section is ${\cal O}(\alpha)$ suppressed,
but one might enjoy a longer run on the $\Upsilon(5S)$ for reasons
of $B_s$ physics. One could also gain in $1/E_\gamma$ enhancement
when $Y_b$ is closer to $\Upsilon(5S)$, but the narrow width
approximation may start to be questionable. Since we do not know
the width for $Y_b$, we just use Table~1 as a rough guide. With 24
fb$^{-1}$ on $\Upsilon(5S)$, assuming $\Gamma_{Y_b\to ee}{\cal
B}_{Y_b\to \pi^+\pi^-\Upsilon}$ is similar Eq.~(1), even for
$M_{Y_b}\sim 10600$ MeV one expects close to 600
$\pi^+\pi^-\ell^+\ell^-$ events, where $\ell = e,\;\mu$ and
$m_{\ell\ell}$ reconstruct to $M_\Upsilon$. $Y_b$ mass closer to
$\Upsilon(5S)$ would give more events. Thus, even for
$\Gamma_{Y_b\to ee}{\cal B}_{Y_b\to \pi^+\pi^-\Upsilon}$ as low as
1 eV, one can get similar significance for $Y_b$ as the BaBar
discovery of $Y(4260)$, where 125 events were obtained from 211
fb$^{-1}$ data on the $\Upsilon(4S)$. It seems that ISR return on
$\Upsilon(5S)$ would definitely find the corresponding $Y_b$ if it
is lighter than 10865 MeV in mass.

The ISR return from $\Upsilon(5S)$ does not cover the full range
of lattice predictions for the $b\bar bg$ hybrid. One may have to
directly scan for $e^+e^- \to Y_b \to \pi^+\pi^-\Upsilon$ if $Y_b$
turns out to be heavier than $\Upsilon(5S)$. Let us estimate the
cross section involved. This is a standard resonance cross
section, hence
\begin{eqnarray}
\sigma_0(s)
 \simeq \frac{12\pi{\cal B}_{ee}{\cal B}_{\pi^+\pi^-\Upsilon}}{s}
 \sim \frac{1027}{M^2_{Y_b}}\;{\rm pb} \sim 9 \; {\rm pb},
 \nonumber
\end{eqnarray}
where $s = M_{Y_b}^2$ is in GeV$^2$ units, and we have taken
${\cal B}_{ee}{\cal B}_{ \pi^+\pi^-\Upsilon}$ to be the same as
for $Y_c(4260)$. The cross sections for the three mass values are
given in Table~1, but can be easily extended above as it varies
slowly.

With just 0.013 fb$^{-1}$ of data on the $Y_c(4260)$, CLEO-c was
able to observe\cite{CLEOc} a clean signal of 37 $\pi^+\pi^-J/\psi
\to \pi^+\pi^-\ell^+\ell^-$ events with little background. If Eq.
(1) holds approximately for $Y_b \to \pi^+\pi^-\Upsilon$, what is
the integrated luminosity needed to achieve the same for $Y_b$?
The $1/s$ drop in cross section costs a factor of 0.16, while
$\Gamma_{\ell\ell}$ costs a factor of 0.4. With background and
trigger issues, the efficiency could be lower. Thus, even if
Eq.~(1) holds for the $Y_b$, one would probably need $\sim 0.3$
fb$^{-1}$ data or more to reach a similar number of events as the
CLEO-c scan for $Y_c$. After all, CLEO-c knew the target mass.
Without knowing more precisely the $Y_b$ mass range, it seems
difficult to pursue a scan search over a wide range.
Maybe at the future Super B factory, the issue could be more
easily covered.

\section{Discussion and Conclusion}

Let us offer a few remarks.

Only narrow states are accessible at hadronic machines, so $X_b$
should be pursued, and higher energy machines have the advantage.
But heavy ion environment would be too difficult because of pion
combinatorics. Since $1^{++}$ and many other quantum numbers are
difficult to access at $e^+e^-$ machines, LHCb should be prepared
for other narrow states as well. Also, LHCb has the capability to
explore a more diverse program, such as $K^+K^-\Upsilon$,
$\omega\Upsilon$ etc., as well as J$^{\rm PC}$ determination.
Thus, LHCb may well enjoy the bonanza in bottomonium spectroscopy
that the B factories have enjoyed for charmonium.

For $Y_b$, Belle should exploit their $\Upsilon(5S)$ data for ISR
return search. It would probably be a difficult decision for BaBar
to make to run on $\Upsilon(5S)$ to compete. However, at 0.3
fb$^{-1}$ per scan point, the scan search for $Y_b$ above the
$\Upsilon(5S)$ may have to await the end of the B factory era, the
Super B factory era, or a strong stimulus that focuses the target
mass range. On the other hand, if the $Y_b$ is found by ISR return
study on $\Upsilon(5S)$, a dedicated direct production study would
be much profitable to learn about more detailed properties, such
as $K^+K^-\Upsilon$ and $\pi^0\pi^0\Upsilon$ decay channels.

In conclusion, we think that the bottom counterparts of $X(3872)$
and $Y(4260)$, called the $X_b$ and $Y_b$, could possibly be
discovered in the near future.

\section*{Acknowledgments}
I thank Alex Bondar for a critical remark on the scanning approach
for $Y_b$.

\balance

\appendix

\end{document}